# Comparison of the Phenomena of Light Refraction and Gravitational Bending

Robert J. Buenker, Bergishe Universitaet Wuppertal,

*Fachbereich C-Theoretische Chemie, Gaussstr. 20, D-42119 Wuppertal, Germany*

## Abstract

The properties of light in the presence of electromagnetic and gravitational fields are compared.  Once one takes account of the fact that clock rates vary with distance from a massive object, it is argued that in an absolute sense light frequencies remain constant in both interactions.  It is also pointed out that the criterion used by Einstein for the angle of curvature of light rays passing close to the sun is not their actual trajectory but rather Huygens' Principle.  The latter only requires that the speed of light vary with distance from a gravitational source in order to produce a measurable effect. As a result, the observed displacement of star images during solar eclipses can be explained on the basis of a rotation of the wave front of light without assuming that individual photons are actually deflected by the sun.  A calculation reported by Schiff in 1960 based on the assumption that light travels in a straight line for all local observers obtains the same closed expression for the angle of displacement of star images as in Einstein's original work, in support of this interpretation.  Since light is believed to follow a straight-line trajectory within any given homogeneous transparent medium, it is argued that light refraction and gravitational bending have more in common than is generally realized.

## I. Introduction

In recent work [1,2] it has been shown that the quantum mechanical formulas $E=\hbar\omega$ and $p=\hbar k$ can be used to at least partially reconcile Newton's corpuscular theory of light with experiment.  The main conclusion in *Opticks* [3] was that light is bent when entering a transparent medium from free space because there is a force acting normal to the interface. As a consequence, the component of momentum parallel thereto must be conserved in accordance with the Second Law of Kinematics.  In modern terminology this observation means that the momentum p of the photons (corpuscles of light) is proportional to the refractive index n of the medium in which they travel.  Since n is greater for water than for free space (n=1), Newton was led to conclude that *both* the momentum and the speed v of the photons increase as they enter the denser medium. When the speed of light was measured in water in 1850 by Foucault, it is well known that Newton's prediction was found to be incorrect, and on this basis Newton's *Opticks*, including most especially his particle theory of light, fell into disrepute within the scientific community [4].

Consistent application of the above quantum mechanical formulas [1,2] shows something quite different, however, when Hamilton's canonical equations are used to compute the light speed v, namely as the derivative of the energy E with respect to the momentum p [5]. When this is done, the observed dependence of v on n is obtained. In particular, it is found that, in quantitative agreement with experiment, the speed of light decreases as it passes from free space into water or any other normally dispersive medium. Newton's conclusion on the basis of his Second Law, namely that the momentum p of the photons increases as they enter a dense medium from air, forms the theoretical basis for this result. Therefore, the Second Law is verified by the light speed measurements rather than being contradicted by them, and as a result one of the most widely used arguments against the particle theory of light is shown to be specious.

The objective of the present study is to compare the phenomenon of light refraction, which is caused by the electromagnetic interaction between the photons and the molecules of the transparent medium, with the gravitational bending of light [6,7]. The latter is clearly a much weaker effect which can only be observed by means of careful measurements of the apparent positions of celestial bodies [8], whereas light refraction can easily be observed with the naked eye. Yet one knows that both forces are proportional to the inverse square of the distance between interacting objects, and this similarity has led to a longstanding search for a unification of the electromagnetic and gravitational interactions [9,10], thus far without success. The most obvious distinction between these two forces is the fact that gravitation is only observed to be attractive, whereas electrically charged objects can both be attracted and repelled as a result of the Coulomb interaction. There has been speculation that antimatter might be repelled in gravitational fields, but there are at least strong theoretical arguments against this view [11]. The knowledge that standard quantum mechanical formulas can be used to better understand the phenomenon of light refraction suggests that their application to gravitational bending may also prove to be instructive, particularly with regard to the ongoing search for a theory of quantum gravity, and this line of approach will be considered in the following discussion.

## II. Light Frequency Variation and Gravitational Scaling of Units

One of the clearest distinctions between light refraction and the phenomenon of gravitational bending of light can be seen from the effect each has on the frequency of light ω. In refraction the frequency does not change as the light passes from one medium to another. By contrast, one has the gravitational red shift [12,13] when light is emitted from the sun or other stars. Accordingly, the measured frequency of a given atomic line varies with the position of both the observer and the light source in a gravitational field. For an observer on earth the measured frequency decreases the closer the light source is to the sun, for example.

The other major distinction is that in refraction the light changes its direction at the interface between two transparent media but otherwise travels in a straight line, whereas the gravitational bending is thought to occur continuously as the light passes by a massive object. Because of Newton's First and Second Laws it can safely be assumed that in the case of refraction there is only a net force acting on the light at the interface along the direction of the normal. Inside the medium there are no unbalanced forces to alter the direction of the photons. There is thus a constant potential throughout a given medium, whereby there is general agreement that the interaction itself is electromagnetic in nature. The potential simply changes from one constant value to another as the light passes through an interface between two different transparent media.

In gravitational bending it is assumed instead that light is bent continuously as it passes a massive object such as the sun. The classical (Newtonian) theory of gravitation assumes that the field is conservative and therefore that the total energy is constant along the entire path for a given object, but that potential and kinetic energies are constantly being interchanged. Einstein [12] had a different explanation for this phenomenon, however. He concluded that the unit of energy changes as the distance from a gravitational source varies. When light travels from the sun to the earth, its energy E is the same along the entire path, but the numerical value measured by a series of observers constantly decreases in inverse proportion to the energy unit as it moves to a higher potential.

He also argued that the same variation with distance from the gravitational source occurs for light frequency, which result is consistent with Planck's radiation law,

$$E = h\,\hbar\,\omega. \qquad (1)$$

In other words, on an absolute basis the frequency of the light is actually the same on the sun as it is when it is measured upon arriving at the earth. Since the unit of frequency is smaller near the sun, however, its numerical value is found to be lower on earth than if the same atomic line were generated in a laboratory there. This effect has been verified by observing the blue shift of iron-57 x-rays as they fall through a distance of only 20 m [15]. When viewed in this way, there is actually no fundamental difference between the two cases when light traverses a gravitational field or passes between different transparent media. Both the energy E and the frequency ω of the photons are perfectly conserved at every step along the way for both phenomena.

# III. Light Deflection

A. Huygens' Principle

Given the latter similarity for light frequencies, it is instructive to consider the aforementioned distinction in the trajectories of light as it passes between different transparent media or traverses a gravitational field. In this case it is important to recall how the angle $d\theta$ of gravitational bending is defined in Einstein's original work [14,16]. Huygens' Principle is used, which in its differential form gives

$$d\theta = 1/c' \ (dc'/dy) \ dx, \qquad (2)$$

where $c'$ is the speed of light, y is the lateral distance from a gravitational source such as the sun, and dx is the distance traveled by the light in a given time interval. The key consideration in determining the magnitude of the angle $d\theta$ is that the value of $c'$ decreases as the light draws nearer to the sun [14]. The local speed of light is always c, the standard value of $2.997 \times 10^8$ m/s, but because of the changes in the units of time and distance with gravitational potential [14, 16], the value measured on earth ($c'$) depends on the position of the source.

It is important to see that the criterion of eq. (2) does not necessarily imply that light actually moves along a curved path. On the contrary, the diagram in Fig. 1 shows that two light rays traveling close to the sun along perfectly straight-line trajectories will lead to a nonzero value for $d\theta$. This is because the value of $c'$ is known to vary with the lateral distance from the sun (y). As a result the outer light ray will travel a greater distance in a given time interval dt than does the inner one. The difference is dc' dt, that is, the differential in light speed multiplied by the elapsed time. To a satisfactory approximation, this time interval is related to the distance traveled by the light dx divided by the average speed of light, so that dt = dx/c'. The line connecting the two end points of the light trajectories therefore *rotates* relative to the initial orientation of the two light beams by an angle $d\theta$ which is given by the ratio of dc'dx/c' to dy, the lateral distance differential for the two light beams. The result is eq. (2), the same equation as Einstein used in his original work [14,16] to predict the angle of displacement of star images during solar eclipses.

There is a straightforward interpretation of the latter result. When light enters the human eye or some photographic device, the direction from which it comes is judged by extending the normal to the associated wave front backward in space. When the speed of light varies with location of the light beam, it is entirely possible that this method of judging the direction from which it has come produces what amounts to an optical illusion. The illusion comes about *not because the trajectory of the light is actually bent by the sun*, however, but rather because its *wave front is rotated relative to its original orientation*. The distinction is significant in the present context. It shows that the behavior of light in refraction and in the presence of a gravitational source may have much more in common than is generally thought to be the case.

## B. Schiff's Method

To explore this point further it is interesting to consider the shifting of star images during solar eclipses in a more quantitative fashion. Einstein's prediction of the amount of the shift was based on the general theory of relativity [16]. Schiff reported a simpler method many years later [17], however, which relied exclusively on Einstein's earlier work [14] on the scaling of the units of time and distance under the influence of gravitational fields. He showed that if one integrates $d\theta$ in eq. (2) over the entire path of the light from infinity to the earth's surface, the result is in quantitative agreement with Einstein's value for the total angle of displacement of the star images. He obtained the same closed expression as Einstein for this quantity, including the dependence on both the gravitational mass of the sun or other object and the distance of closest approach of the light to it.

The key assumption in Schiff's method is *that light always travels in a straight line with speed c for a local observer*, that is, one who is at the same gravitational potential as the light. On this basis it is a simple matter to compute the instantaneous speed of light for an observer on earth using Einstein's techniques of gravitational scaling. By substitution into eq. (2) and then integrating, one obtains the desired result for the total amount of angular displacement of star images. It is important to note that no trajectory for the light beam is actually calculated in this procedure. Schiff does state that "since c' increases with increasing y, the curvature is such that the ray is concave toward the sun," in agreement with the experimental finding that stars images are displaced outward from the sun during eclipses. Einstein came to the same conclusion much earlier [16]. Since c' is known at each value along the path of the light ray, it is possible to actually plot the light trajectory as it passes by the sun, however. It will be shown below that the result is not at all what is implied in the above statement.

Let us start with the case of the light approaching the sun from above (see Fig. 1). First of all, the gravitational scaling of time implies that the speed of light observed on earth is less than c. This is because the clocks on earth are running faster than they are close to the sun. According to Huygens' Principle in eq. (2), this means that the wave front of the light is rotated because dc'/dy is not zero. Changing the measure of time cannot affect the direction that individual objects move, however. This means that the *light must continue to move in the same direction as before, that is, in a straight line*. There is thus already an indication that there is a distinction between the Huygens' Principle criterion for the "bending" of light and that which one normally uses for objects based on their trajectories.

In 1916 Einstein realized that it was also necessary to scale the unit of radial distance [16]. This development was key to his successful prediction of the angle of displacement of star images during solar eclipses. Failure to scale the radial distance as well leads to exactly half the correct value [14, 17]. Since this scaling is not isotropic it becomes possible for the theoretical treatment to actually find a trajectory for the light that is not a straight line. On closer examination, however, one finds that the *change indicated is actually in the wrong direction*. Because the meter stick on earth is larger than near the sun, it follows that the radial component of the light velocity is observed to be smaller on earth than for a local observer close to the sun. Since the tangential component of the light velocity is the same for both observers, it follows that the light must appear to veer *away* from the sun for the observer on earth. In other words, the light path becomes curved when computed on this basis, but the direction of the curvature is convex, the opposite of what is known from actual measurements of star image displacements. This tendency continues until the light reaches a position that is closest to the midpoint of the sun. Afterwards, the opposite tendency is predicted, because now decreasing the radial component causes the light to veer toward the sun relative to the straight-line trajectory of the local observer. The result is a Δ-shaped trajectory for the light. It bears no resemblance to the uniformly concave trajectories commonly found in the literature. Moreover, the light is still moving along the same straight line when it reaches the earth as when it left the star, since the aforementioned deflection is only indicated when the light is in the immediate vicinity of the sun.

One is left with the conclusion that the above way of predicting the actual trajectory of the light is therefore flawed, and that Einstein's correct prediction of the amount of the "bending" of light is based *solely* on Huygens' Principle. As shown above, however, it is entirely possible to obtain this value by assuming that light actually follows the straight-line trajectory indicated in Fig. 1. All that's really necessary is that c' vary with lateral distance from the sun, and this is taken care of quantitatively by employing Einstein's scaling procedure and assuming that adjacent rays of light are moving in perfectly straight lines along the entire journey from the star to the earth.

## C. Soldner's Calculation and Ascoli Scaling of Acceleration

The idea that light travels along a curved trajectory in the presence of massive objects did not originate with Einstein, but rather goes back at least to the time of Newton. Soldner [7] carried out a calculation based on Newton's universal gravitation theory in the early nineteenth century.  He assumed that the light particles were subject to the same g field as any other object, consistent with Galileo's unicity principle, and obtained a curved trajectory as a result.  The angle of deflection of the light beam was almost exactly half of Einstein's later value.  It is very close to what one obtains if the scaling of radial distance is neglected in Einstein's relativistic theory [6,12,17].   The path is clearly concave because of the attractive nature of the gravitational force in Newtonian physics.

   The agreement with Einstein's 1911 result for the angle of deflection (i.e.,  half the correct value) is purely coincidental, however, because the methods used to obtain the respective values are so fundamentally different.  In particular, the criterion used to define the angle of displacement of star images is not the same in the two approaches, as discussed above.   In Soldner's calculation, the trajectory of the light rays must be curved, whereas in Einstein's theory all that is necessary is for the speed of light to vary with lateral distance from the sun.  The question that arises is whether there is any independent justification for Schiff's assumption that, at least for a local observer, light moves in a straight line with constant speed.

    It is helpful to rephrase the question by asking why the acceleration due to gravity is zero for a photon, as Schiff [17] has assumed for each local observer.  Ascoli [18] approached the matter on a general basis by considering how the acceleration of an object varies from one inertial system to another.   He concluded that if an object moving with relative speed v is subject to a local acceleration $a_L$, the observer must measure a smaller acceleration $a_O$ such that

$$a_O = a_L \gamma^{-2} (v), \qquad (3)$$

where $\gamma (v) = (1- v^2/c^2)^{-0.5}$.   Since light moves with speed c relative to the observer and $a_L= g$, the local acceleration due to gravity, it follows from this equation that $a_O=0$, that is, that the observer must measure a constant value for the speed of light.  The derivation that Ascoli gave is based on Einstein's velocity addition law [18] and has the disadvantage that it only holds if the velocity and acceleration vectors are pointed in the same direction.  It is possible to overcome this deficiency, however, by noting that clocks moving with the accelerated object run more slowly than those of the observer by a factor of $\gamma (v)$ [19].  Since acceleration is second-order in time, this means that $a_L$ must be $\gamma^2 (v)$ times larger than $a_O$.   One assumes thereby that an observer moving with the object must measure all distances to be the same as for the primary observer.  The reason is that relativistic contraction is not physical and so the moving observer has no way of detecting any changes in the dimensions of his environment (isometric corollary to the relativity principle [19]).

   In summary, there are good theoretical reasons for believing that light follows a straight-line trajectory in free space, no matter how strong the gravitational field in a given region.   Soldner's classical approach to the calculation of the light trajectory [7] is deficient because it fails to take into account that clock rates vary with gravitational potential. Relativistic treatments [16,17] obtain the correct angle of displacement of star images by gravitational sources, but they do so only when Huygens' Principle is used to define this quantity. When one tries to calculate the actual trajectory of light on this basis, the result is contradictory. One obtains a Δ-shaped curve that is partially convex to the sun. This result indicates that the methods employed by Einstein [16] and Schiff [17] succeed in calculating the speed of light c' from the vantage point of an observer on earth, but they give a misleading impression of the actual trajectory followed by the light on its way from infinity past the sun.

## IV. Black Holes and Light Refraction

One of the most interesting conjectures about the gravitational bending of light is that the effect might be great enough to prevent the escape of light from extremely dense objects. The first mention of this idea appears to have been made by Mitchell [20] in the eighteenth century. Soldner also discussed the idea of "black holes" in his work [7]. The arguments of the last section indicate something quite different, however, namely that light isn't actually deflected by massive bodies. Rather the wave front of light is rotated by them in such a way as to make it appear that the light source is located farther away from the massive body than is actually the case. As indicated in Fig. 1, two light rays grazing the surface of the sun travel at different speeds. According to Huygens' Principle, the magnitude of the effect depends on the rate of change of the speed of light with lateral distance from the massive object. The amount of rotation will increase with the gravitational mass of the object, but the trajectory of the light will remain a straight line in any case.

Again a clear similarity exists between this phenomenon and light refraction. Some materials will bend light more than others but in all cases the trajectory of the individual light rays will be a straight line once they have entered a given medium. A change in the angle of incidence relative to that of refraction occurs when the electromagnetic properties of the medium change at an interface. In gravitational interactions there is no analog for an interface, however. As long as light is traveling through free space, there is nothing to cause it to alter its direction. The closer the light beam is to the gravitational source, however, the more it slows down from the vantage point of an observer located on the surface of the earth or at any other fixed gravitational potential. Because of Huygens' Principle this means that the wave front of the light appears to rotate, creating the illusion that the source of the light has changed its position relative to the case when the massive body is not present.

As long as the refracting medium is transparent, it can never prevent light from escaping it. The indication from the above discussion is that the same is true in gravitational "bending." There is experimental evidence for the existence of black holes, but the only thing that would prevent light from escaping them is if all the individual photons were absorbed on their way to the surface. Light rays that just graze the outer surface of the black hole will simply continue on in a straight line and thus should be readily observable. The associated wave front of the light will be strongly rotated, however. If the light source is large enough so that light rays can be observed on either side of the black hole, the phenomenon of "gravitational lensing" can occur. It becomes impossible to obtain a sharp image of the light source because wave fronts passing on opposite sides of it are rotated in opposite directions.

Finally, it should be noted that the above discussion also suggests a novel way to measure the speed of light in refractive media. When light enters the medium at an angle $\phi$, the wave front also rotates (see Fig. 2). As a consequence, the image of the light source appears to be in a different position when viewed from inside the medium than from without. The normal to the wave front inside the medium makes a different angle ($\phi$") with the interface than the angle of refraction ($\phi$'). Measurement of both angles allows one to obtain the ratio of the group refractive index $n_g$ to the normal refractive index n. By definition, the speed of light inside the refractive medium is $c/n_g$. Conventionally, $n_g$ is obtained by measuring n for various wavelengths of light, whereas the present approach allows for its determination by making different angle measurements for the same light source. The underlying principle is seen to be the same as in Fig. 1. Adjacent light rays travel at different speeds for a certain interval as they enter the dispersive medium, and this causes the wave front to rotate so that the image of the light source appears to have been displaced. Observation of this effect in transparent media would constitute a verification of the present interpretation for gravitational displacement of star images, particularly the conclusion that observation of the latter phenomenon is completely consistent with straight-line trajectories for light in the neighborhood of massive bodies.

## V.  The Hypothesis of a Null Gravitional Potential

Newton's First Law states that a body will continue moving in a straight line if there are no unbalanced forces acting upon it. When applied to light refraction it indicates that the electromagnetic potential within a homogeneous transparent medium is constant throughout. The fact that light rays change direction at an interface between two such media simply implies that the value of the constant potential is not the same in both. Once it is realized that gravitational bending can be explained consistently by assuming that light always travels in straight lines in free space but at different speeds depending on its proximity to a massive object, the possibility presents itself that the gravitational potential is also constant. Moreover, since the trajectory of light is not changed as it moves toward the observer from infinity, it seems clear that the value of this constant must be zero.

One of the consequences of this hypothesis is that the quantum mechanical Hamiltonian must be the same for all observers independent of their position in a gravitational field. This statement is at least consistent with Einstein's assumption that local observers must agree on the values of atomic frequencies [14]. In his biography of Einstein, Pais [21] states that such a conclusion cannot be rigorously correct, since "we do know that such a displacement must exist," that is, one that is caused by a local gravitational field. This is a question that can only be solved experimentally, however. One has to show that a given atomic line has a measurably different frequency/energy when it is located at a different gravitational potential than is present in laboratories on the surface of the earth. In the absence of such determinations, the hypothesis of a null potential due to gravity remains viable.

Such a hypothesis makes quantum gravity extremely simple. It means that the local Hamiltonian is the same for everyone, regardless of his location in a gravitational field. Differences in properties of atoms and molecules measured by observers at different gravitational potentials can always be reconciled by taking into account the effects of gravitational scaling for length, time and mass. The rates of chemical reactions are the same for all local observers once the effects of temperature and pressure are taken into account. No additional correction needs to be made for gravity itself. In essence this hypothesis belies the old adage that "gravity cannot be painted onto the rest of physics." The same quantum mechanical equations are valid everywhere in the universe. One simply has to take account of the effects of gravitational scaling of units to predict the results of experiments carried out in different parts of the universe.

## VI. Conclusion

The properties of light in transparent media and in gravitational fields have been compared and found to have many common features. To begin with, energy is conserved in both phenomena. Einstein changed the viewpoint of physicists in the early twentieth century by showing that changes in the energy of a fallen object can be more consistently explained in terms of a change in units rather than an exchange of kinetic and potential energy. He extended this argument to include light frequencies and time. On the one hand, this led him to predict a gravitational red shift, but on the other, it allowed one to argue on absolute terms that the frequency of light emitted near the sun is exactly the same for a local observer as for his counterpart in a laboratory on earth. The numerical value actually measured is different, but this is only because the unit of frequency becomes smaller as the distance from a gravitational source decreases. In light refraction the frequency is observed to be the same in each transparent medium, but this is because the unit does not vary in this case.

The present study has extended these arguments to deal with the question of whether light is really deflected by gravitational fields.  One knows that light rays are bent at the interface between different media, but inside a given medium they are observed to travel in straight lines.  The commonly held view of gravitational bending is that light follows a continuously curved trajectory in the presence of massive bodies such as the sun, so that on this basis the analogy with light refraction appears to break down.  Closer inspection finds, however, that the criterion on which Einstein based his successful prediction of the gravitational displacement of star images has nothing directly to do with the actual trajectory followed by light on its way to the earth.   Instead, the angle of displacement is obtained from Huygens' Principle, which only relies on information about the variation of the speed of light with lateral distance from the gravitational source.   The diagram in Fig. 1 shows that it is entirely possible to obtain a displacement of star images based on this criterion without having the light follow a curved trajectory.   Gravitational scaling is known to cause the speed of light to decrease the closer it comes to the sun.  As a consequence the *wave front of the light rotates* away from the massive object, creating the illusion that the stars themselves have changed their position.

When one considers how this variation in the speed of light comes about in gravitational theory, the case for a straight-line trajectory gains in credibility.  Schiff demonstrated that it is possible to calculate the angle of displacement by assuming that for an observer at the same gravitational potential the speed of light   has the same constant value of c and travels in a perfectly straight line.   Einstein's procedures for the gravitational scaling of the units of time and distance then allow one to correctly predict the variation of light speed c' measured by an observer on earth, which in turn leads to the same closed expression for the displacement angle as a function of the mass of the gravitational source and distance of closest approach of the light as is obtained in the general theory of relativity.   Half of the computed angle comes from the scaling of time, however, which can only result in a change in the speed of light observed on earth, not its direction.  The implication is thus that at least this part of the effect is still derived on the basis of a straight-line trajectory.   Moreover, scaling of the radial component of the velocity implies a light trajectory that veers away from the sun on its initial approach, thereby producing a curvature for the light path that is convex, not concave as is commonly assumed.   The conclusion is that while the gravitational scaling correctly predicts the speed of light, it is incapable of telling us anything about its trajectory.   This again leaves open the distinct possibility that the light travels in a perfectly straight line for all observers, not just one located at the same gravitational potential.

There is an additional theoretical argument in favor of the assumption of a straight-line trajectory for light in the presence of massive bodies.   If one takes account of the variation of clock rates from one inertial system to another, it follows that the acceleration experienced by an object in relative motion to the observer will be measured by him to be smaller by a factor of $(1-v^2/c^2)$.  This result, which was first obtained by Ascoli [see eq. (3)], implies that the acceleration due to gravity for light has a null value for any observer.  At the very least, this result is consistent with Schiff's assumption that the locally observed velocity of light is constant in both speed and direction.  It also proves to be possible to compute the other key quantity in the general theory of relativity, the advancement angle of the perihelion of planetary orbits, by using Einstein's gravitational scaling of units and Ascoli's relativistic damping of acceleration, as will be discussed in subsequent work [22].   The procedure is quite simple and can be carried out with a Fortran computer program of approximately 100 statements [23] that is applicable to both the planetary orbit and the star image displacement calculations.

   The main point that needs to be emphasized in the present discussion is that the well-known observation of star image displacements during solar eclipses does not constitute experimental proof that light follows a curved trajectory when passing by massive bodies.  Just as in light refraction, the observed effect can be explained satisfactorily in terms of a rotation of the wave front of the light formed by rays that travel in perfectly straight lines.

## Figure Captions

Fig. 1. Schematic diagram showing light rays emitted by stars to follow straight-line trajectories as they pass near the sun. Because of gravitational effects the speed of the light rays c' is known to increase with gravitational potential, with the effect that the corresponding Huygens wave front gradually rotates away from the sun. As discussed in the text, the normal to a given wave front points out the direction from which the light appears to have come, causing the star images to be displaced by an angle $\vartheta$ during solar eclipses.

Fig. 2. Angles of incidence $\Phi$ and refraction $\Phi'$ as light enters water from free space. The corresponding angle $\Phi''$ which the wave front makes with the interface upon entering a dispersive medium can be determined by noting the direction from which the image of the light source appears to arrive inside the medium. Note that $\Phi''$ is generally different than $\Phi'$.

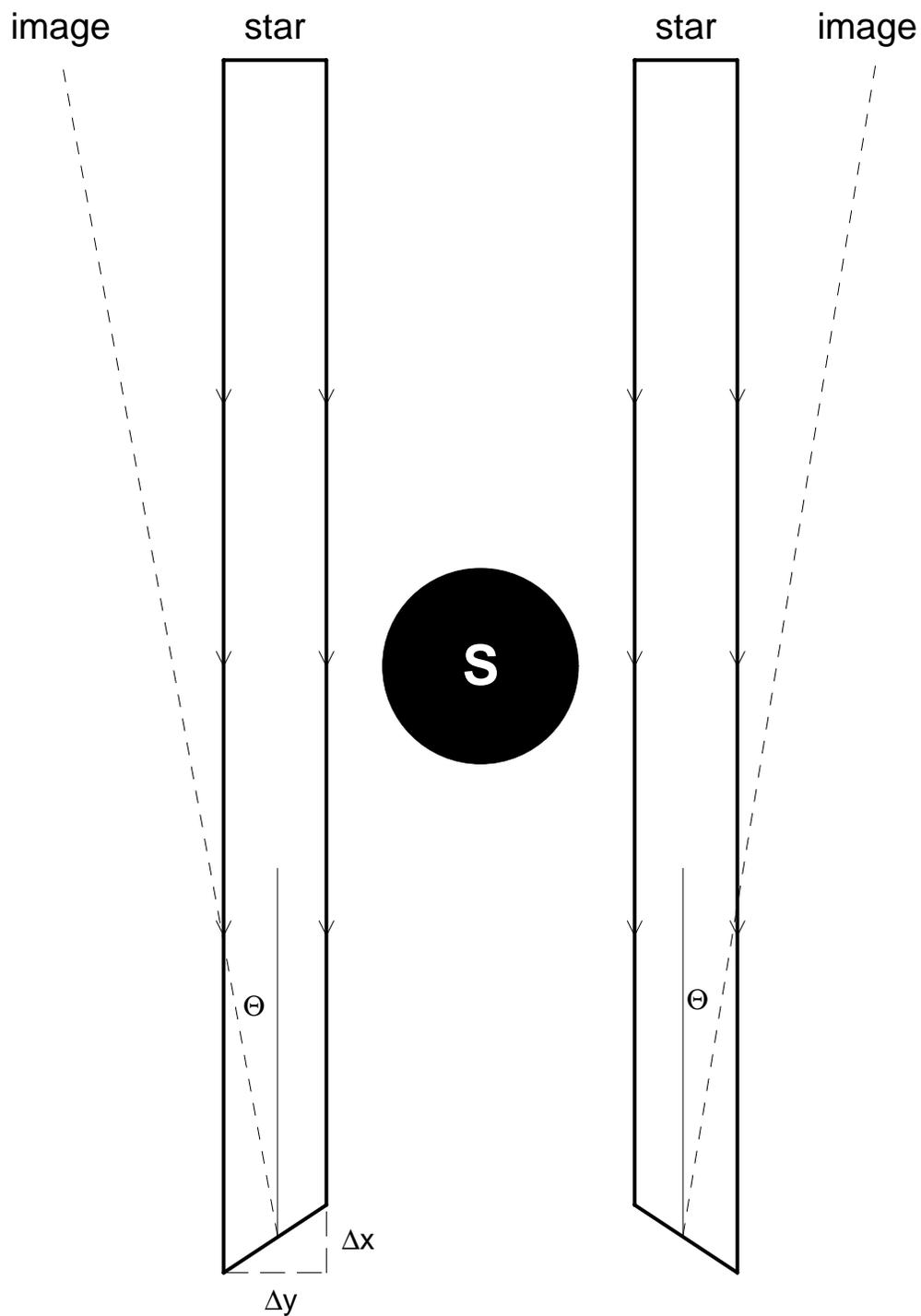

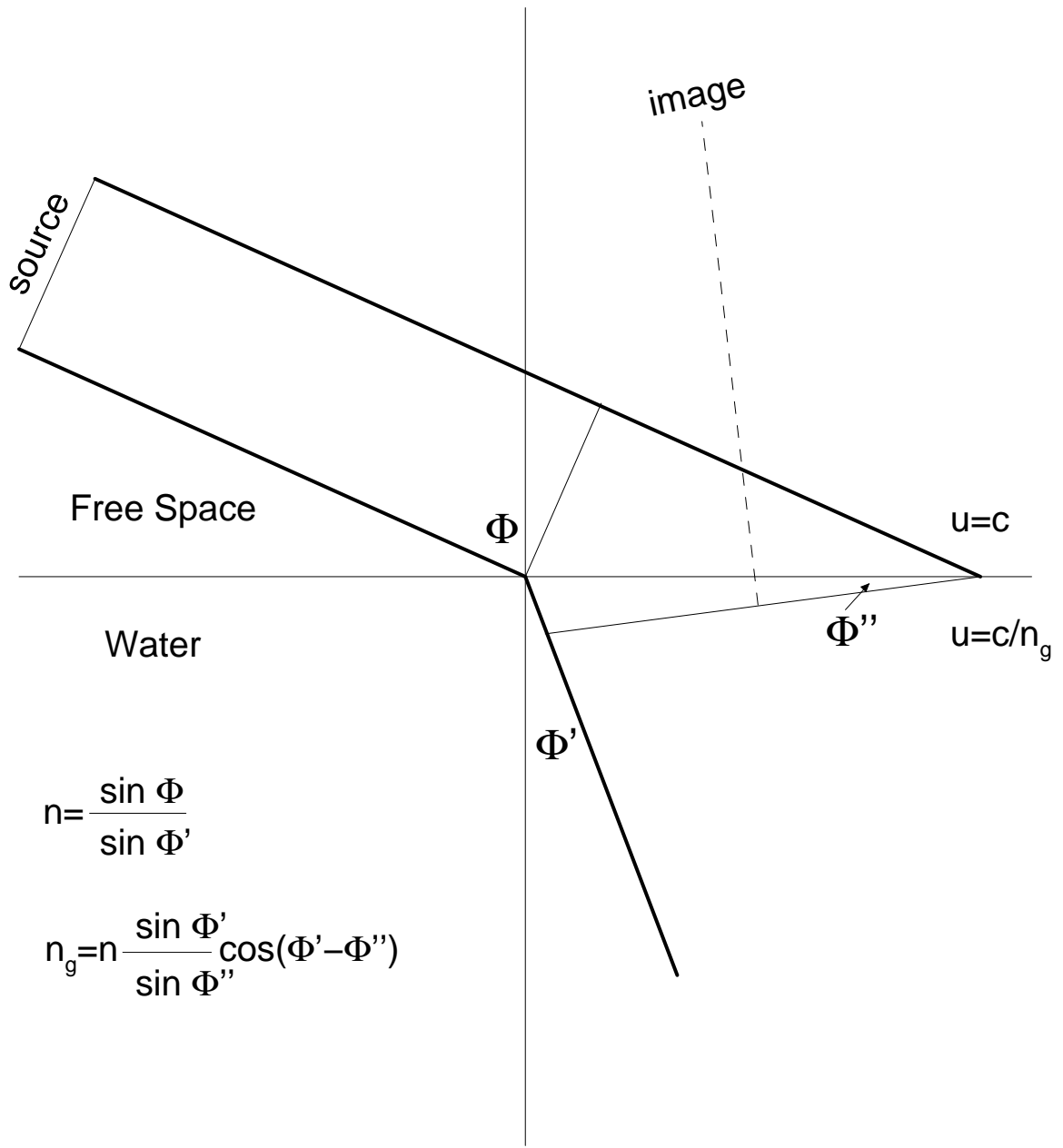